\documentclass[journal]{IEEEtran}

\usepackage{algorithm}
\usepackage{algorithmic}
\usepackage{amsmath,amssymb}
\usepackage{bm}
\usepackage{amsfonts}
\usepackage{amsthm}
\usepackage{multirow}
\usepackage{graphicx}
\usepackage{graphics}
\usepackage{color}
\usepackage{url}
\usepackage{array}
\usepackage{verbatim}

\usepackage{booktabs}

\begin{document}
%



\title{LipSound2: Self-Supervised Pre-Training \\for Lip-to-Speech Reconstruction and Lip Reading}

%
%
%


\author{Leyuan Qu, Cornelius Weber and Stefan Wermter 
\thanks{L. Qu, C. Weber and S. Wermter are with the Knowledge Technology Group, Department of Informatics, University of Hamburg, Vogt-Kölln-Straße 30, 22527 Hamburg, Germany. E-mail: qu@studium.uni-hamburg.de, \{weber, wermter\}@uni-hamburg.de}}

\markboth{IEEE Transactions on Neural Networks and Learning Systems}%
{Shell \MakeLowercase{\textit{et al.}}: Bare Demo of IEEEtran.cls for IEEE Journals}


\maketitle

\begin{abstract}

The aim of this work is to investigate the impact of crossmodal self-supervised pre-training for speech reconstruction (video-to-audio) by leveraging the natural co-occurrence of audio and visual streams in videos. We propose LipSound2 which consists of an encoder-decoder architecture and location-aware attention mechanism to map face image sequences to mel-scale spectrograms directly without requiring any human annotations. The proposed LipSound2 model is firstly pre-trained on $\sim$2400h multi-lingual (e.g. English and German) audio-visual data (VoxCeleb2). To verify the generalizability of the proposed method, we then fine-tune the pre-trained model on domain-specific datasets (GRID, TCD-TIMIT) for English speech reconstruction and achieve a significant improvement on speech quality and intelligibility compared to previous approaches in speaker-dependent and -independent settings. In addition to English, we conduct Chinese speech reconstruction on the CMLR dataset to verify the impact on transferability. Lastly, we train the cascaded lip reading (video-to-text) system by fine-tuning the generated audios on a pre-trained speech recognition system and achieve state-of-the-art performance on both English and Chinese benchmark datasets.









\end{abstract}







%
\IEEEpeerreviewmaketitle

\section{Introduction}
\label{section:Introduction}


\IEEEPARstart{I}{nspired} by human bimodal perception~\cite{besle2004bimodal} in which both sight and sound are used to improve the comprehension of speech, a lot of effort has been spent on speech processing tasks by leveraging visual information, for example, integrating simultaneous lip movement sequences into speech recognition~\cite{chung2017lip, afouras2018deep}, guiding neural networks in isolating target speech signals with a static face image for speech separation~\cite{qu2020multimodal,chung2020facefilter} and grounding speech recognition with visual objects and scene information~\cite{miao2016open,gupta2017visual}. Multi-modal audio-visual methods achieve significant improvement over single modality models, since the visual signals are invariant to acoustic noise and complementary to auditory representations~\cite{macdonald1978visual}. Moreover, the visual contribution becomes more important as the acoustic signal-to-noise ratio is decreased~\cite{silsbee1996computer}.

In most approaches, the visual information is mainly used as auxiliary input to complement audio signals. However,  in some circumstances, the auditory information may be absent or extremely noisy, which motivates speech reconstruction. Speech reconstruction aims to generate both intelligible and qualified speech by only conditioning on image sequences of talking mouths or faces. Generating intelligible speech from silent videos enables many applications, e.g. a silent visual input method on mobile phones for privacy protection in public areas~\cite{denby2010silent}; communication assistance for patients suffering laryngectomy~\cite{sharifzadeh2010reconstruction}; surveillance video understanding when only visual signals are available~\cite{cristani2007audio}; enhancement of video conferences or far-field human-robot interaction scenarios in a noisy environment~\cite{tsiami2018far}; non-disruptive user intervention for autonomous vehicles~\cite{tscharn2017stop}. 


It is challenging to reconstruct qualified and intelligible speech from only mouth or face movements, since human speech is produced by not only externally observable organs, like lips and tongue, but also internally invisible ones which are difficult to capture in most cases~\cite{gick2012articulatory}, for instance, vocal cords and pharynx. Consequently, it is hard to infer fundamental frequency or voicing information controlled by these organs. Moreover, some phonemes are acoustically discriminative but not easy to distinguish visually since the phonemes share the same places of articulation but with different manners of articulation~\cite{maeda1990compensatory}, for example, /v/ and /f/ in English are both fricatives and look the same on lip and teeth movements but are different on the vibration of vocal cords (voiced vs unvoiced) and the attribute of aspirate (unaspirated vs aspirated) which are not visible in most video recordings. Hence, predicting human voices from appearance is still a challenging task~\cite{goto2020face2speech}.

In recent years, there has been a growing interest in speech reconstruction and variant methods have been proposed. A possible technique is to run lip reading (video-to-text) and text-to-speech (TTS) systems in cascade but the lip reading performance is still unsatisfactory and the error is being propagated to TTS. Alternatively, other researchers directly estimate speech representations, for example, linear predictive coding~\cite{ephrat2017vid2speech}, bottleneck features~\cite{akbari2018lip2audspec}, and mel-scale spectrograms~\cite{qu2019lipsound},
from videos, followed by a vocoder used to transform intermediate representations to audio, for instance, STRAIGHT~\cite{kawahara1999restructuring} and WORLD vocoder~\cite{morise2016world}. In contrast, the information of speaker identity and speaking styles can be relatively preserved. However, most existing work only focuses on speaker-dependent settings with a small vocabulary or artificial grammar dataset, or even builds one model for each individual speaker, which does not meet the requirements in realistic scenarios. 


\begin{figure}[t]
  \setlength{\abovecaptionskip}{0pt}
  \setlength{\belowcaptionskip}{1pt}
  \centering
  \includegraphics[width=\linewidth]{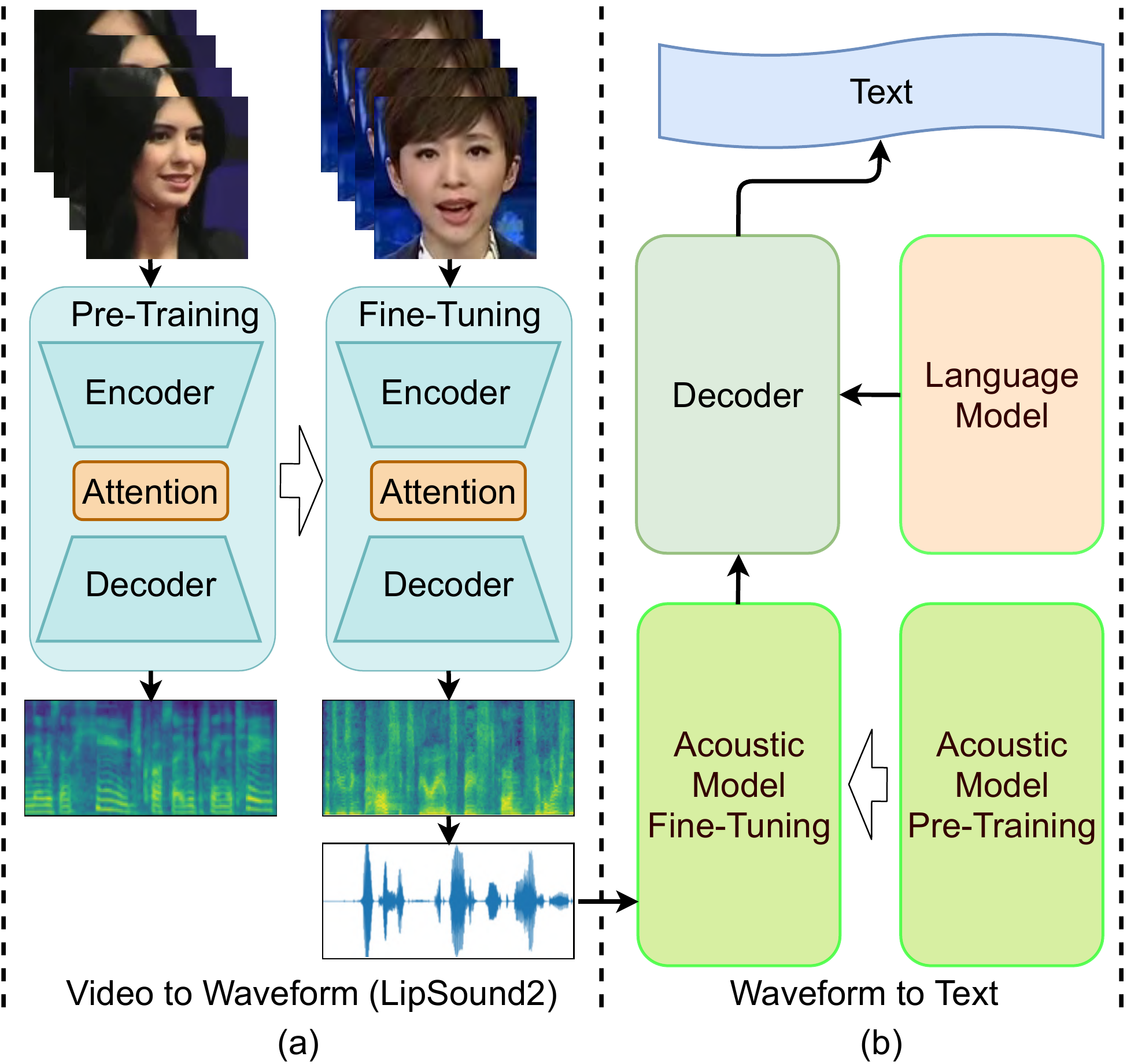}
  \caption{Process of video-to-waveform generation and waveform-to-text transformation.}
  \label{fig:pipeline}
  \vspace{-0.5cm}
\end{figure}

In our previous work, we proposed LipSound~\cite{qu2019lipsound} to directly map visual sequences to low-level speech representation, i.e. mel-spectrogram, which is inspired by audio-visual self-supervised representation learning. By leveraging the natural co-occurrence of audio and visual streams in videos without requiring any human annotations, or treating one modality as the supervision of the other, self-supervised representation learning has received substantial interest, for example, learning representations by matching the temporal synchronization~\cite{korbar2018cooperative} or spatial alignment~\cite{morgado2020learning} of audio and video clips for action recognition. 

In comparison to our previous work LipSound that only focuses on speaker-dependent settings for the GRID artificial grammar dataset, in this paper, we further explore to what extent the large scale crossmodal self-supervised pre-training can benefit speech reconstruction in generalizability (speak-independent) and transferability (Non-Chinese to Chinese) on a large vocabulary continuous speech corpus TCD-TIMIT. In addition, we also changed the LipSound architecture substantially by replacing 1DCNN with 3DCNN blocks (Conv 3D + Batch Norm + ReLU + Max Pooling + Dropout). This should enable the model to directly learn stable representations from raw pixels and using location-aware attention mechanism to make the alignments between encoder and decoder more robust to non-verbal areas. Moreover, we replace the Griffin-Lim algorithm~\cite{griffin1984signal}  with a neural vocoder to smoothly generate waveforms and voices.

As shown in Fig.~\ref{fig:pipeline} (a), our approach is first pre-training the Lipsound2 model on a large-scale multi-lingual audio-visual corpus (VoxCeleb2) to map silent videos to mel-spectrogram, then fine-tuning the pre-trained model on specific domain datasets (GRID, TCD-TIMIT and CMLR), followed by a neural vocoder (WaveGlow~\cite{prenger2019waveglow}) to reconstruct estimated mel-spectrogram to waveforms. Lip reading (video-to-text) experiments are performed by fine-tuning the generated audios on a pre-trained acoustic model (Jasper~\cite{li2019jasper}) in Fig.~\ref{fig:pipeline} (b).

The main contributions of this paper are: 
\begin{enumerate}
\item We propose an auto-regressive encoder-decoder with attention architecture, LipSound2, to directly map silent facial movement sequences to mel-scale spectrograms for speech reconstruction, which does not require any human annotations.
\item We explore the model generalizability on speaker-independent and large-scale vocabulary datasets which few studies have focused on, and we achieve better performance on speech quality and intelligibility in the speech reconstruction task.
\item To the best of our knowledge, no previous research has investigated Chinese speech reconstruction in speaker-dependent and -independent cases.
\item By leveraging the large-scale self-supervised pre-training on LipSound2 and the advanced Jasper speech recognition model, our cascaded lip reading system outperforms existing models by a margin on both English and Chinese corpora.
\end{enumerate}

The paper is organised as follows. Section~\ref{section:related work} reviews related work on lip-to-speech reconstruction, lip reading and self-supervised learning. Section~\ref{section: model architecture} provides the model details, followed by the description of datasets and evaluation metrics in Section~\ref{section: Experimental Setup}. Experimental results and discussion are presented in Section~\ref{section: Results and Discussion} and Section~\ref{discussion} respectively. We conclude this paper in Section~\ref{conclusion}.

\section{Related Work}
\label{section:related work}
\subsection{Lip to Speech Reconstruction}

In recent years, researchers have investigated a variety of approaches to speech reconstruction from silent videos. We only review the neural network methods in this paper.

Le Cornu et al.~\cite{cornu2015reconstructing} propose to use fully connected neural networks to estimate spectral envelope representations, for instance linear predictive coding (LPC) coefficients and Mel-filterbank amplitudes, from visual feature inputs, such as two-dimensional discrete cosine transform, followed by a STRAIGHT vocoder ~\cite{kawahara1999restructuring}, which is used to synthesize time-domain speech signals from the estimated representations. Follow-up work~\cite{le2017generating} predicts speech-related codebook entries with a classification framework to get further improvement on speech intelligibility. Instead of using handcrafted visual features, Ephrat et al.~\cite{ephrat2017vid2speech} utilize convolutional neural networks (CNNs) to automatically learn optimal features from raw pixels and show promising results on out-of-vocabulary experiments. Subsequently, improved results are reported by Ephrat et al.~\cite{ephrat2017improved} via combining RestNet backbone and a post-processing network on a large-scale vocabulary dataset, TCD-TIMIT~\cite{harte2015tcd}. Akbari et al.~\cite{akbari2018lip2audspec} treat the intermediate bottleneck features learned by a speech auto-encoder as training targets by conditioning on lip reading network outputs. Kumar et al. ~\cite{kumar2019lipper} validate the effectiveness of using multiple views of faces on both speaker-dependent and -independent speech reconstruction. Vougioukas et al. ~\cite{vougioukas2019video} utilize generative adversarial networks (GAN) to directly predict raw waveforms from visual inputs in an end-to-end fashion without generating an intermediate representation of audios. Inspired by the speech synthesis model, Tacotron2~\cite{shen2018natural}, Qu et al. propose to directly map video inputs to low-level speech representations, mel-spectrogram, with an encoder-decoder architecture and achieve better results on lip reading experiments. Afterwards, Prajwal et al. ~\cite{prajwal2020learning} improve the model performance with 3D CNN and skip connections. Recently, Michelsanti et al. ~\cite{michelsanti2020vocoder} have presented a multi-task architecture to learn spectral
envelope, aperiodic parameters and fundamental frequency separately, which are then fed into a vocoder for waveform synthesis. They integrate a connectionist temporal classification (CTC)~\cite{graves2006connectionist} loss to jointly perform lip reading, which is capable of further enhancing and constraining the video encoder. 

In addition to sequences of lip or face images, further signals can be used for temporal self-supervision. For instance, Gonzalez et al.~\cite{gonzalez2017direct} generate speech from articulatory sensor data and Akbari et al.~\cite{akbari2019towards} reconstruct speech from invasive electrocorticography. However, most existing work only focuses on a speaker-dependent setting and small vocabulary or artificial grammar datasets. In this paper, we evaluate our method not only on speaker-dependent experiments but also pay attention to speaker-independent and large-scale vocabulary setups.



\subsection{Lip Reading}

Lip reading, also known as visual speech recognition, is the task to predict text transcriptions from silent videos, such as mouth or face movement sequences. Research on lip reading has a long tradition. 
Approaches to lip reading generally fall into two categories on feature level: a) handcrafted visual feature extraction, such as  Discrete Cosine Transform~\cite{heckmann2002dct}, Discrete Wavelet Transform~\cite{potamianos1998image} or Active Appearance Models~\cite{sterpu2018towards}; b) representations learned by neural networks, which has become the dominant technique for this task, for example, using convolutional auto-encoders~\cite{parekh2018lip}, spatio-temporal convolutional neural networks \cite{assael2016lipnet}, long short-term memory~\cite{wand2016lipreading}, or residual networks \cite{stafylakis2017combining}.

Alternatively, methods on modeling units for lip reading can be divided into word- and character-level. a) In the case of word-level units, lip reading is simplified as a classification task. Word-level lip reading datasets and benchmarks are built, for instance LRW~\cite{chung2016lip} for English and LRW-1000~\cite{yang2019lrw} for Chinese. Stafylakis et al.~\cite{stafylakis2017combining} adopt spatiotemporal convolutional networks and 2D ResNet as front end to extract visual features and bidirectional Long Short-Term Memory networks as the backend to capture temporal information, and attain significant improvement. Weng et al. ~\cite{weng2019learning} present two separated deep 3D CNN front ends to learn features from grayscale video and optical flow inputs, respectively. Martinez et al.~\cite{martinez2020lipreading} replace recurrent neural networks widely used in past work with Temporal Convolutional Networks to simplify the training procedure. The word-level methods are usually able to achieve high accuracy, however, the models disregard the interaction or co-articulation phenomenon between phonemes or words. A predefined lexicon with closed-set vocabulary is used and words are usually treated as isolated units in speech. Thereby, long-term context information, assimilation or dissimilation effects are completely neglected. Moreover, it is hard to recognize out-of-vocabulary words.

b) Lip reading models with character- or phoneme-level mainly use methods proposed in speech recognition. Assael et al.~\cite{assael2016lipnet} conduct end-to-end lip reading experiments on sentence-level with CTC loss. Subsequently, sequence discriminative training~\cite{thangthai2017improving} and domain-adversarial training~\cite{wand2017improving} are introduced to lip reading.  Chung et al. ~\cite{chung2017lip} collected the dataset, ‘Lip Reading Sentences’ (LRS) which consists of hundreds of thousands of videos from BBC television, and significantly promote the research on sentence-level lip reading. Shillingford et al.~\cite{shillingford2018large} verify the effectiveness of large-scale data (3,886 hours of video) for training continuous visual speech recognition. Afouras et al.~\cite{afouras2018deeplip} compare the performance of recurrent neural networks, fully convolutional neural networks and Transformer on lip reading character recognition. 

Different from the mainstream methods which directly transform videos to text, we perform lip reading experiments in a cascaded manner, in which the silent videos are firstly mapped to audios with our LipSound2 model, then text transcriptions are predicted by fine-tuning on a pretrained speech recognition system. 



\vspace{-0.1cm}
\subsection{Self-supervised Learning}

As a form of unsupervised learning, self-supervised learning leverages massive unlabelled data and aims to learn effective intermediate representations with the supervision of self-generated labels. Training unlabelled data in a supervised manner relies on the pretext tasks that determines what labels and loss functions to be used. In computer vision, the pretext tasks can be predicting angles of rotated images ~\cite{gidaris2018unsupervised},  learning the relative position of segmented regions in an image ~\cite{doersch2015unsupervised}, placing shuffled patches back~\cite{noroozi2016unsupervised} or colorizing grayscale input images~\cite{zhang2016colorful}. The video-based pretext tasks can be tracking moving objects in videos~\cite{wang2015unsupervised}, validating temporal frame orders~\cite{misra2016shuffle} and video colorization~\cite{vondrick2018tracking}, and so on. 

Self-supervised learning is also widely used in natural language processing. It has made substantial progress recently, where diverse pretext tasks are proposed, for instance, predicting center words using surrounding ones or vice versa~\cite{mikolov2013efficient}, generating the next word by conditioning on previous words in an auto-regressive fashion~\cite{radford2018improving}, completing masked tokens or consecutive utterances~\cite{devlin2018bert}, recovering the order of shuffled words~\cite{lan2019albert} or the permutation of rotated sentence~\cite{lewis2019bart}. 


Inspired by the strong correlation between different modalities where, for example, the audio and visual modalities are consistent semantically or happen synchronously, more and more researchers work on multi-modality or cross-modality self-supervised learning. Multi-modality self-supervised learning aims to learn joint or shared latent spaces or representations while cross-modality self-supervised learning perceives one modality as the supervision of the other. Here we only review the audio-visual modalities since this is the main focus of our paper. Different pretext tasks are designed according to the correspondence and synchronization of audio and visual modalities, for instance, predicting whether image and audio clips correspond to enable neural networks to classify sounds~\cite{arandjelovic2017look}, learning cross-modal retrieval~\cite{chung2019perfect}, locating the sound source in an image~\cite{arandjelovic2018objects}, learning representations by matching the temporal synchronization~\cite{korbar2018cooperative} or spatial alignment~\cite{morgado2020learning} of audio and video clips for action recognition, combining a contrastive loss and a clustering loss to learn high-level semantic representations for visual events and concepts understanding~\cite{chen2021multimodal}. In this paper, we focus on cross-modal self-supervised learning where the corresponding audio signals are treated as the supervisions of face sequence inputs.




\begin{figure*}[!t]
\setlength{\abovecaptionskip}{0cm}
\setlength{\belowcaptionskip}{-0cm}
\centering 
\small
\includegraphics[width=16cm]{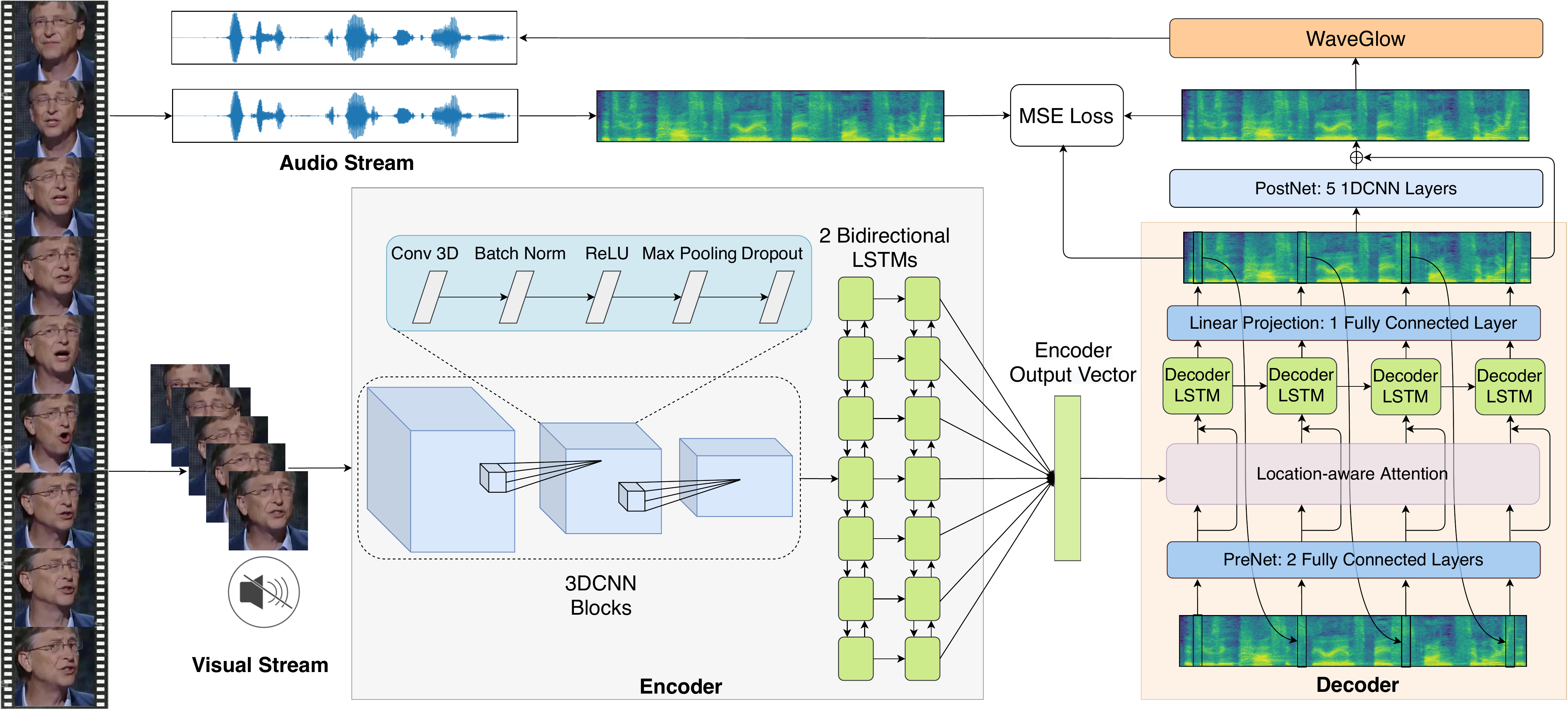}
\caption{\small The architecture of LipSound2. The video is split into visual and acoustic streams. The face region which is cropped from the silent visual stream is used as the model input. The acoustic spectrogram features extracted from the counterpart audio stream are used as the training target. During training, the ground truth spectrogram frames are utilized to accelerate convergence, while, during inference, the outputs from previous steps are used.}
\label{architecture}
\vspace{-0.5cm}
\end{figure*}


\section{Model Architecture}
\label{section: model architecture}


Fig.~\ref{architecture} shows the LipSound2 model architecture. We split the video clips into an audio stream used as training target and a visual stream used as model input. The system consumes the visual part to predict the audio counterpart in a self-supervised fashion. The proposed architecture is composed of an encoder-decoder and an attention model to map the soundless visual sequences to the low-level acoustic representation, mel-scale spectrograms. Advantages are that, in contrast to directly predicting raw waveform, working with mel-spectrogram not only reduces computational complexity but also easily learns long-distance dependence. Model details are listed in Table~\ref{tab:model_cofiguration}. Then a pre-trained neural vocoder, WaveGlow, follows to reconstruct raw waveform from the generated mel-spectrogram. 

\subsection{Encoder}

The multi-Task CNN (MTCNN) \cite{zhang2016joint} is used to detect face landmarks from raw videos. We crop only the face region ($112 \times 112 $ pixels) and smooth all frame landmarks, since low-resolution videos or profile faces lead to detection failures sometimes and landmark smoothing can eliminate frame skip in adjacent images. The cropped face sequences are then fed into 3D CNN blocks and each block is based on a 3D CNN, Batch Normalization, ReLU activation, Max Pooling and Dropout, as shown in Fig.~\ref{architecture}. Then two bidirectional LSTM layers follow which capture the long-distance dependence from the left and right context.

\subsection{Location-sensitive Attention}

\begin{figure}[H]
  \setlength{\abovecaptionskip}{0cm}
  \setlength{\belowcaptionskip}{-1pt}
  \centering
  \includegraphics[width=8cm]{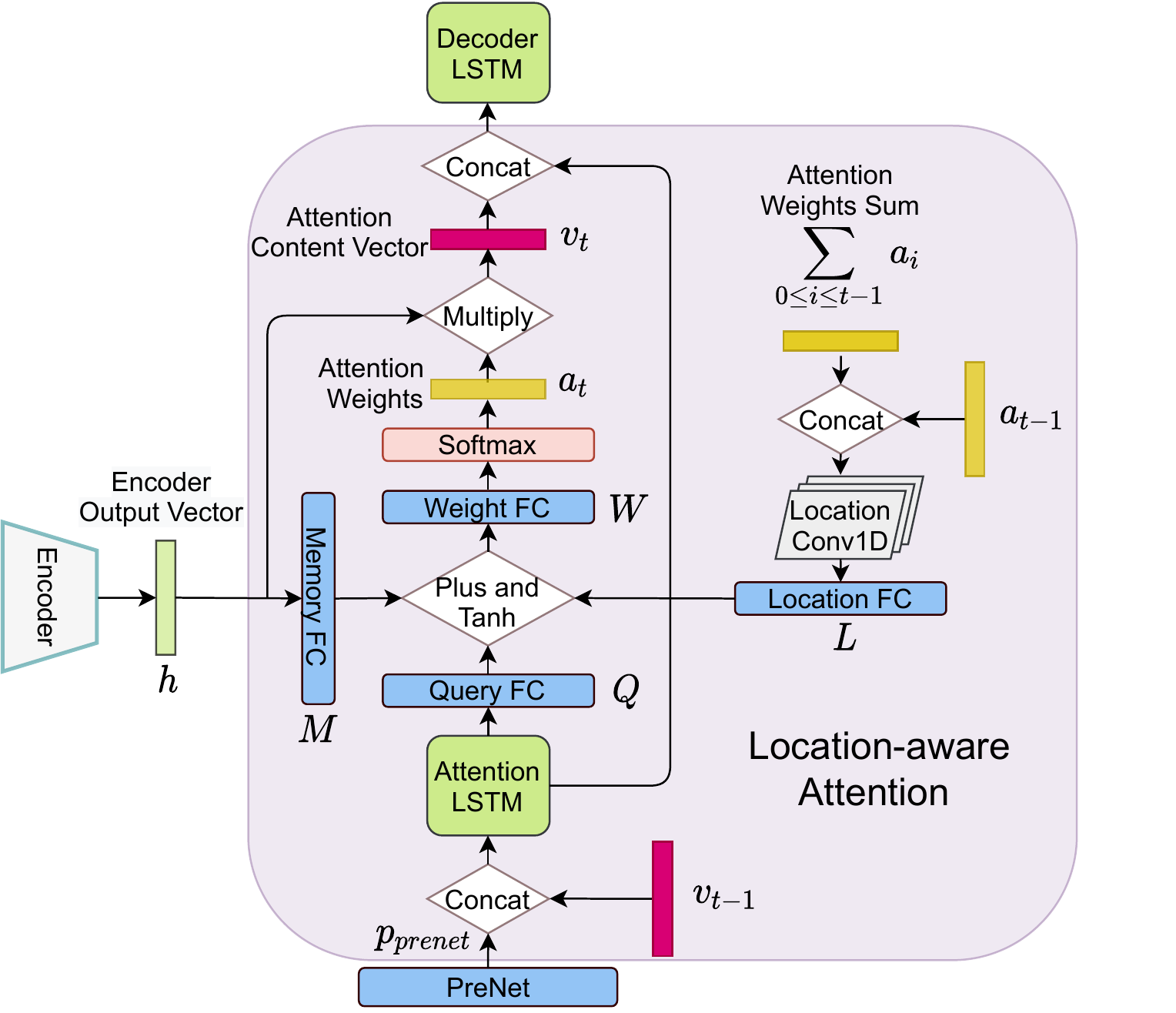}
  \caption{The computational flow of location-aware attention at time step $t$.}
  \label{fig:attention}
\end{figure}

We use location-aware attention~\cite{chorowski2015attention} to bridge the encoder and the decoder. The image sequence input $i = (i_{0},...,i_{n})$ is firstly embedded into the latent space representation vector $h = (h_{1},...,h_{n})$ by the encoder with the same dimension $n$ in time, then the intermediate vector $h$ is decoded into the mel-spectrogram $o = (o_{0},...,o_{m})$. At time step $t\ (0\le t\le m)$, the attention weight $a_{t}$ can be obtained by the following equations:

\begin{equation}
\setlength{\abovedisplayskip}{1pt}
\setlength{\belowdisplayskip}{1pt}
a_{t}=Softmax(W \cdot tanh(M \cdot h+Q \cdot x+L \cdot y))
\label{attention1}
\end{equation}

\begin{equation}
\setlength{\abovedisplayskip}{1pt}
\setlength{\belowdisplayskip}{1pt}
x=LSTM(h \cdot a_{t-1}, p_{prenet})
\label{attention2}
\end{equation}

\begin{equation}
\setlength{\abovedisplayskip}{1pt}
\setlength{\belowdisplayskip}{1pt}
y=Conv(a_{t-1}, \sum_{0\le i\le t-1}{a_{i}})
\label{attention3}
\end{equation}

where $W, M, Q, L$ are the matrices learned by Weight FC (fully connected), Memory FC, Query FC and Location FC, respectively. In Eq.~\eqref{attention3}, the sum of attention weights of all previous steps is integrated, which enables the current step attention to be aware of the global location and move forward monotonically. Fig.~\ref{fig:attention} visualizes the computational flow of the attention mechanism. The attention content vector $v_{t}$ can be obtained by multiplying the encoder output by the normalized attention weights (Eq.~\eqref{attention4}).

\begin{equation}
\setlength{\abovedisplayskip}{1pt}
\setlength{\belowdisplayskip}{1pt}
v_{t}=a_{t} \cdot h
\label{attention4}
\end{equation}




\subsection{Decoder}

The decoder module consists of one unidirectional LSTM layer and one linear projection layer. The decoder LSTM consumes the attention content vector and the output from attention LSTM to generate one frame at a time. Subsequently, the linear projection layer maps the decoder LSTM outputs to the dimension of the mel-scale filter bank. During training, we use ground truth mel-spectrogram frames as PreNet inputs and during inference, the predicted frames from previous time steps are used. Since the decoder only receives past information at every time step, after decoding, five Conv1D layers (postnet) are used to further improve the model performance by smoothing the transition of adjacent frames and using future information which is not available when decoding.

\begin{table}[th]
  \setlength{\abovecaptionskip}{0cm}
  \setlength{\belowcaptionskip}{-0cm}
  \caption{Configuration of LipSound2 encoder, decoder, attention and postnet.}
  \label{tab:model_cofiguration}
  \centering
  \setlength\tabcolsep{2pt}%
  \resizebox{9cm}{!}{
    \begin{tabular}{ccccc}
    \toprule
    \textbf{Layer} & \textbf{Kernel} & \textbf{Stride} & \textbf{Padding} & \textbf{Channels/Nodes}\\
    \midrule
    \textbf{Encoder} \\
    Conv3D 1                   & $5 \times 3 \times 3$         & $[1,2,2] $          & $ [2,0,0] $                      & 32                           \\
    MaxPool3D &$ 1 \times 2 \times 2 $  & $[1,2,2]$ & $[0,0,0]$& - \\ 
    Conv3D 2                   & $ 5 \times 3 \times 3 $            & $[1,2,2]$           & $[2,0,0]$                        & 64                           \\
    MaxPool3D &$ 1 \times 2 \times 2 $ & $[1,2,2]$ & $[0,0,0]$& - \\ 
    Conv3D 3                   & $ 5 \times 3 \times 3 $             & $[1,1,1]$           & $[2,0,0] $                         & 128                           \\
    MaxPool3D &$ 1 \times 2 \times 2 $ & $[1,2,2]$ & $[0,0,0]$& - \\ 
    BiLSTM1                & -           & -           & -                         & 128                           \\
    BiLSTM2                & -           & -           & -                         & 128                           \\
    \midrule
    \textbf{Attention} \\
    Attention LSTM & -           & -           & - & 1024\\
    Query FC & -           & -           & - & 128\\
    Memory FC& -           & -           & - & 128\\
    Location Conv1D             & 31           & 1          & 15  & 32\\
    Location FC& -           & -           & - & 128\\
    Weight FC & -           & -           & - & 1\\
    \midrule
    \textbf{Decoder}\\
    PreNet FC 1 & -           & -           & - & 512\\
    PreNet FC 2 & -           & -           & - & 256\\
    Decoder LSTM & -           & -           & - & 1024\\
    Linear Projection FC& -           & -           & - & 80\\
    \midrule
    \textbf{PostNet} \\
    Conv1D 1                   & 5           & 1          & 2                         & 512                           \\
    Conv1D 2                   & 5           & 1           & 2                         & 512                           \\
    Conv1D 3                   & 5           & 1          & 2                        & 512                           \\
    Conv1D 4                   & 5          & 1           & 2                         & 512                           \\
    Conv1D 5                   & 5           & 1           & 2                         & 80                           \\
    \bottomrule
    \end{tabular}}
\end{table}

\subsection{Training Objective}
The loss function is the sum of two mean square errors (MSE), as shown in Eq. \eqref{loss}, i.e. the MSE between the decoder output $O_{dec}$ and the target mel-spectrogram $M_{tar}$ and the MSE between the postnet output $O_{post}$ and the target mel-spectrogram.

\begin{equation}
\setlength{\abovedisplayskip}{1pt}
\setlength{\belowdisplayskip}{1pt}
Loss=MSE(O_{dec}, M_{tar}) + MSE (O_{post}, M_{tar})
\label{loss}
\end{equation}

\subsection{WaveGlow}

We use WaveGlow~\cite{prenger2019waveglow} which combines the approach of the glow-based generative model~\cite{kingma2018glow} and the architecture insight of WaveNet~\cite{oord2016wavenet} to transform the estimated mel-spectrogram back to audio. WaveGlow abandons auto-regression~\cite{oord2016wavenet} and speeds up the procedure of waveform synthesis in high quality and resolution. We train WaveGlow from scratch using the same settings as original work~\cite{prenger2019waveglow} but in 16k sampling rate on the LJSpeech dataset~\cite{ljspeech17} to meet the requirement of following up ASR models. To our surprise, the WaveGlow model that is trained with only one female voice can effectively generalize to any unseen voices and stably perform waveform reconstruction.

\subsection{Acoustic Model and Language Model}
The Jasper~\cite{li2019jasper} speech recognition system which is a fully convolutional architecture trained with skip connections and CTC loss is adopted to directly predict characters from speech signals. We pretrain the Jasper DR 10x5 model\footnote{https://nvidia.github.io/OpenSeq2Seq/html/speech-recognition.html} on 960h LibriSpeech and 1000h AISHELL-2 corpora, which achieves 3.61\% WER (word error rate) and 10.05\% CER (character error rate) on the development set for English and Chinese, respectively.

Beam search is utilized to decode the output character possibilities from Jasper and a 6-gram KenLM~\cite{heafield2011kenlm} language model\footnote{https://github.com/PaddlePaddle/DeepSpeech} into grammatically and semantically correct words on sentence-level~\cite{wermter1997screen}.

\section{Experimental Setup}
\label{section: Experimental Setup}
\subsection{Dataset}
\begin{figure}[H]
  \setlength{\abovecaptionskip}{0cm}
  \setlength{\belowcaptionskip}{-0cm}
  \centering
  \includegraphics[width=8cm]{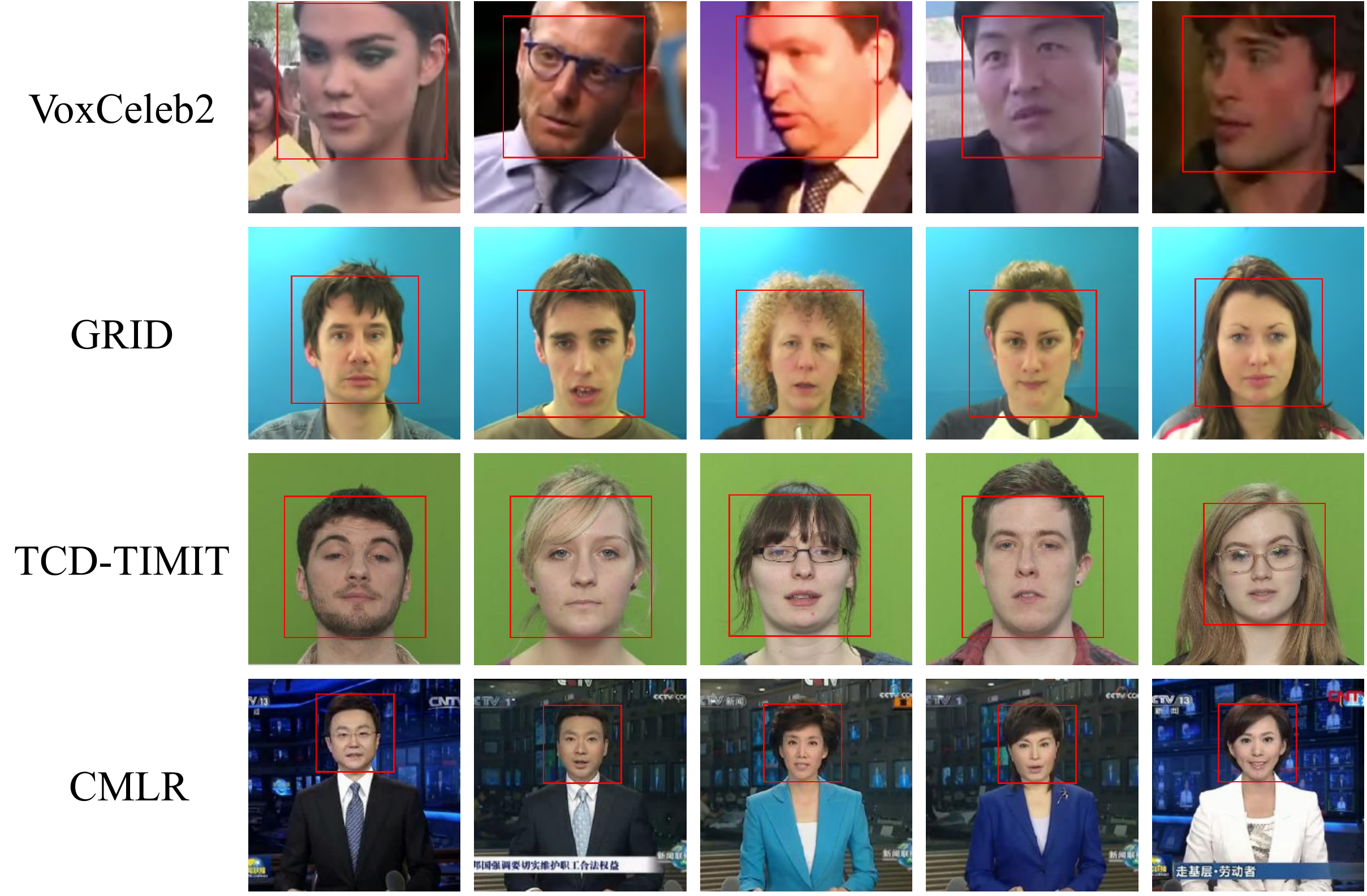}  
  \caption{Random face samples from audio-visual corpora. Only the face region is cropped during training and test.}
  \label{fig:datasets}
  \vspace{-0.5cm}
\end{figure}

All datasets used in this paper are summarized in Table~\ref{tab:datasets} and random frames from audio-visual ones are presented in Fig.~\ref{fig:datasets}. VoxCeleb2 is a large-scale audio-visual corpus, extracted from YouTube videos, containing over one million utterances and more than 6k different speakers from around 145 nationalities and languages. It includes noisy and unconstrained conditions, specifically, the audio stream may be recorded with background noise, such as laughter and room reverberation, and the vision part may contain variable head poses (e.g. frontal faces and profile), variable lighting conditions and low image quality, while the GRID and TCD-TIMIT datasets are in controlled experimental environments with fixed frontal face angle and clean background in audio and vision. It is worth to mention that the GRID dataset is designed to contain only a fixed 6-word structure and all sentences are generated by a restricted artificial grammar: \textit{command + color + preposition + letter + digit + adverb}, for example, set blue in Z three now. CMLR (Chinese Mandarin Lip Reading) is collected from videos by 11 hosts of the Chinese national news program \textit{News Broadcast}, which contains frontal faces and covers a large amount of Chinese vocabulary. We firstly pretrain LipSound2 on VoxCeleb2, then fine-tune the model on GRID, TCD-TIMIT and CMLR respectively for video to mel-spectrogram reconstruction.

LibriSpeech and AISHELL-2 are the current largest open-source speech corpora and widely-used speech recognition benchmarks for English and Chinese, respectively. LibriSpeech is derived from audiobooks, containing 460h of clean speech and 500h of noisy speech. AISHELL-2 consists of 1000h different domain speech, for instance, voice command and smart home scenario, and includes various accents from different areas of China. We use LibriSpeech and AISHELL-2 to pretrain the Jasper acoustic model to boost the performance of waveform-to-text transformation. The generated speech on GRID, TCD-TIMIT and CMLR is used for further fine-tuning to perform lip reading (video-to-text) experiments.

The LJ Speech dataset with only one female voice is especially designed for speech synthesis tasks, which is used for WaveGlow training, in this paper, to transform mel-spectrogram back to waveforms.

\begin{table*}[!t]
\renewcommand\arraystretch{1.2}
\setlength{\abovecaptionskip}{0.cm}
\setlength{\belowcaptionskip}{-0.cm}
	\caption{\small Overview of all corpora used in this paper. Spk: Speakers. Utt: Utterances. Vocab: Vocabulary.}
	\label{tab:datasets}
	\vspace{0.8mm}
	\centering
	\small
	\setlength\tabcolsep{8pt}
\begin{tabular}{cccccccc}
\toprule
\textbf{Language}        & \textbf{Dataset} & \textbf{\#Spk.} & \textbf{\#Utt.} & \textbf{\#Vocab.} & \textbf{\#hours} & \textbf{Usage}                                                                  & \textbf{Modality}             \\ \hline
Multi-Language           & VoxCeleb2 \cite{chung2018voxceleb2}       & 6112            & 1.1M            & -                 & 2442             & \begin{tabular}[c]{@{}c@{}}LipSound2\\ pre-training\end{tabular}                 & \multirow{3}{*}{Audio-Visual} \\ \cline{1-7}
\multirow{4}{*}{English} & GRID \cite{cooke2006audio}            & 51              & 33k             & 51                & 27.5             & \multirow{2}{*}{\begin{tabular}[c]{@{}c@{}}LipSound2\\ fine-tuning\end{tabular}} &                               \\ \cline{2-6}
                         & TCD-TIMIT \cite{harte2015tcd}       & 59              & 5.4k            & 5.9k              & 7                &                                                                                 &                               \\ \cline{2-8} 
                         & LJSpeech \cite{ljspeech17}        & 1               & 13.1k           & -                 & 24               & \begin{tabular}[c]{@{}c@{}}WaveGlow training\end{tabular}                      & \multirow{2}{*}{Audio}        \\ \cline{2-7}
                         & LibriSpeech  \cite{panayotov2015librispeech}     & 2484            & 292.3k          & -                 & 960              & \begin{tabular}[c]{@{}c@{}}Acoustic model\\pre-training\end{tabular}                                                             &                               \\ \hline
\multirow{2}{*}{Chinese} & CMLR \cite{zhao2019cascade}            & 11              & 102k            & 3.5k              & 87.7             & \begin{tabular}[c]{@{}c@{}}LipSound2\\ fine-tuning\end{tabular}                  & Audio-Visual                  \\ \cline{2-8} 
                         & AISHELL-2 \cite{du2018aishell}        & 1991            & -               & -                 & 1000             & \begin{tabular}[c]{@{}c@{}}Acoustic model\\pre-training\end{tabular}                                                                & Audio                         \\ \bottomrule
\end{tabular}
\end{table*}

\subsection{Evaluation Metrics}

We evaluate the generated speech quality and intelligibility with Perceptual Evaluation of Speech Quality (PESQ)~\cite{rix2001perceptual} and Extended Short-Time Objective Intelligibility (ESTOI)~\cite{jensen2016algorithm} respectively. The speech-to-text results are measured with Word Error Rate (WER) and Character Error Rate (CER), the ratio of error terms, i.e., substitutions, deletions and insertions, to the total number of words/characters in the ground truth sequences.

\subsection{Training}

We only describe the training settings of LipSound2 pre-training, LipSound2 fine-tuning and Jasper acoustic model fine-tuning. More details about Japser$^{1}$ pre-training acoustic model, KenLM$^{2}$ language model and WaveGlow~\footnote{https://github.com/NVIDIA/waveglow} can be found on the open source websites.

\subsubsection{Vision Stream}face landmarks are detected using MTCNN \cite{zhang2016joint} from all video frames and only the face area is cropped and reshaped to size of $112 \times 112$ as inputs. We also add one 'visual period' -- an empty frame with all values of 255 -- at the end of every visual stream to help the decoder stop decoding at the right time. A max decoder step threshold of 1000 is activated to terminate decoding when the decoder fails to capture the 'visual period'.


\subsubsection{Audio Stream}we first divide the raw waveforms by the max value to normalize all audios to $[0,1]$, then extract the magnitude using the Short Time Fourier Transform (STFT) with 1024 frequency bins and a 64ms window size with 16ms stride. The mel-scale spectrograms are obtained by applying an 80 channel mel filter bank to the magnitude, followed by dynamic range clipping with a minimum value of 1e-5 and log dynamic range compression.

\subsubsection{LipSound2 Pre-Training} image horizontal flipping, gradient clipping with a threshold of 1.0, early stopping and scheduled sampling~\cite{bengio2015scheduled} are adopted to avoid overfitting. Linear and convolutional layers are initialized with Xavier ~\cite{glorot2010understanding} and tanh functions respectively. We use the cosine learning rate decay strategy with an initial value of 0.001. Our LipSound2 model has around 100M parameters. The audio and visual sequences are both high dimensional data, so we conduct all experiments on 4 NVIDIA Quadro RTX 6000 GPUs with 24G memory in parallel to enable a big batch size. The entire pre-training procedure took around 25 days.

\subsubsection{Fine-tuning}
Pre-trained LipSound2 is fine-tuned on GRID, TCD-TIMIT and CMLR videos respectively to conduct speech reconstruction experiments. Afterwards, the produced speech for English (GRID and TCD-TIMIT) and Chinese (CMLR) is fine-tuned on the pre-trained English (LibriSpeech) and Chinese (AISHELL-2) acoustic models to perform lip reading tasks with a 10 times smaller learning rate.

\section{Experimental Results}
\label{section: Results and Discussion}
\begin{figure*}[!t]
  \setlength{\abovecaptionskip}{0cm}
  \setlength{\belowcaptionskip}{-0cm}
\centering 
\small
\includegraphics[width=16cm]{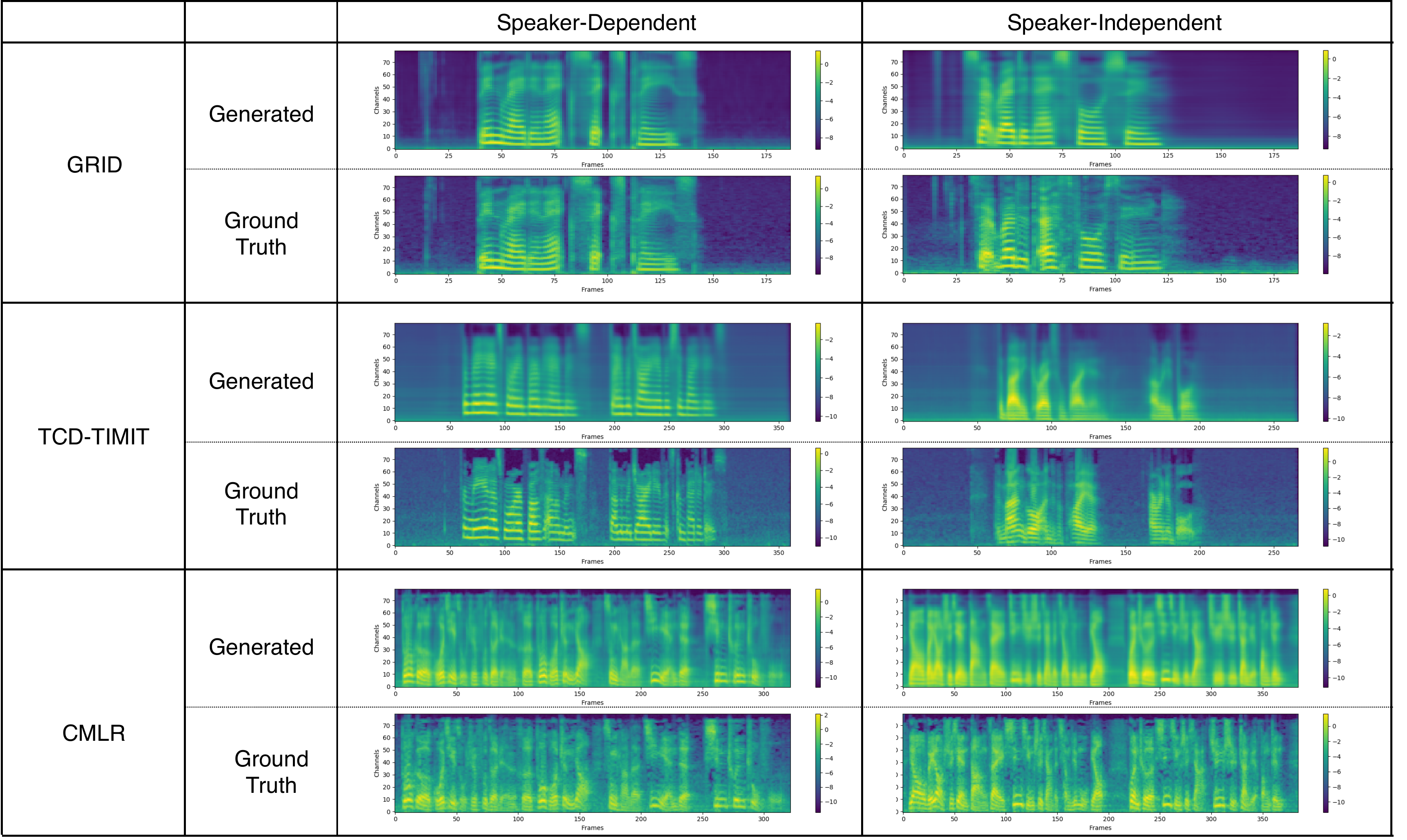}
\caption{\small The comparison between generated mel-spectrogram and ground truth in speaker-dependent and -independent settings for English and Chinese.}
\label{mel}
\vspace{-0.5cm}
\end{figure*}

\subsection{Lip to Speech Reconstruction}
\subsubsection{Speaker-dependent Result}
we report the generated speech results in two perspectives, i.e. speech quality (PESQ) and speech intelligibility (ESTOI). For a fair comparison, we keep the same settings as previous works. 
For speaker-dependent tasks, all datasets are randomly split into 90:5:5 for training, validation and test sets on GRID (Speaker $S1-S4$) and TCD-TIMIT (Lipspeaker $1-3$). Different from previous works that build one model for each individual speaker, we train only one model on all speakers.

As shown in Table~\ref{tab:spk-dep}, our LipSound2 system which is firstly pre-trained on the VoxCeleb2 dataset, then fine-tuned on the specific dataset achieves highest scores on both PESQ and ESTOI, which reveals the effectiveness of our proposed method. The last column in Table ~\ref{tab:spk-dep} compares the number of LipSound2 model parameters against those of baseline systems, showing that its best performance is obtained while staying well in the existing range of numbers of parameters.

\begin{table}[th]
  \setlength{\abovecaptionskip}{0cm}
  \setlength{\belowcaptionskip}{-0cm}
  \setlength\tabcolsep{2pt}
  \caption{Speaker-dependent speech reconstruction results on GRID and TCD-TIMIT datasets.}
  \label{tab:spk-dep}
  \centering
  \resizebox{8cm}{!}{
    \begin{tabular}{cccccc}
    \toprule
                  & \multicolumn{2}{c}{GRID} & \multicolumn{2}{c}{TCD-TIMIT} &  \\ \hline
    Model         & ESTOI        & PESQ       & ESTOI          & PESQ  & Parameters       \\ \hline
    Vid2Speech~\cite{ephrat2017vid2speech}    & 0.335      & 1.734      & 0.298         & 1.136 &   0.9M     \\ 
    Lip2AudSpec~\cite{akbari2018lip2audspec}   & 0.352       & 1.673      & 0.316         & 1.254  &  13.0M     \\ 
    Vougioukas et al.~\cite{vougioukas2019video}     & 0.361       & 1.684      & 0.321         & 1.218 &  not available     \\ 
    Ephrat et al.~\cite{ephrat2017improved} & 0.376       & 1.825      & 0.310         & 1.231 & 9.2M       \\ 
    Lip2Wav~\cite{prajwal2020learning}       & 0.535       & 1.772      & 0.365         & 1.350  & not available      \\
    vid2voc-M-VSR~\cite{michelsanti2020vocoder}       & 0.455       & 1.900     & -         & -   &  5.1M    \\ \hline
    LipSound2          &\textbf{0.592}        &      \textbf{2.328}   & \textbf{0.372} & \textbf{1.490}  &  8.5M  \\\bottomrule
    \end{tabular}
    }
\end{table}

\subsubsection{Speaker-independent Result}
for speaker-independent cases, we follow the same setups for GRID~\cite{vougioukas2019video} and TCD-TIMIT~\cite{harte2015tcd}.

LipSound2 achieves the best results on both metrics on the GRID dataset. Moreover, by listening to the reconstructed audios, we find that our model is capable of producing similar voices as ground truth speakers, instead of generating a weird voice or one of the voices in the training set as occurring in previous works. The model has implicitly learnt the mapping between voices and faces. We highly recommend readers to listen to the produced samples on our demo website\footnote{https://leyuanqu.github.io/LipSound2/}.

Furthermore, we find substitution errors occurring on segment-level (vowels and consonants) because the context information is still not sufficient to disambiguate the phonemes that share the same visible organs, like lips and tongue, but are different in the invisible ones.

To the best of our knowledge, we are the first to tackle the speaker-independent case on the TCD-TIMIT dataset, since TCD-TIMIT consists of limited samples ($\sim$370) for each speaker but with large-scale vocabulary ($\sim$5.9K), which makes the tasks on TCD-TIMIT quite challenging. The speaker-independent results reported in Table~\ref{tab:spk-indep} show considerable performance, for example, the PESQ result is even better than some results reported on speaker-dependent settings (as shown in Table ~\ref{tab:spk-dep}), which suggests that the large-scale self-supervised pre-training enables the model to successfully generalize to unseen speakers.

\vspace{-0.5cm}

\begin{table}[th]
  \setlength{\abovecaptionskip}{0.1cm}
  \setlength{\belowcaptionskip}{-0.1cm}
  \setlength\tabcolsep{2pt}
  \caption{Speaker-independent speech reconstruction results on GRID and TCD-TIMIT datasets.}
  \label{tab:spk-indep}
  \centering
  \resizebox{8cm}{!}{
    \begin{tabular}{cccccc}
    \toprule
                  & \multicolumn{2}{c}{GRID} & \multicolumn{2}{c}{TCD-TIMIT} &  \\ \hline
    Model         & ESTOI        & PESQ       & ESTOI          & PESQ & Parameters         \\ \hline
    Vougioukas et al.~\cite{vougioukas2019video}     & 0.198       & 1.24     & -   & - & not available \\
    vid2voc-M-VSR~\cite{michelsanti2020vocoder}    & 0.227      & 1.23      & -         & - & 5.1M        \\ 
    vid2voc-F-VSR~\cite{michelsanti2020vocoder}    & 0.210      & 1.25      & -         & -  & 5.2M       \\  \hline
    LipSound2          &\textbf{0.363}        &     \textbf{1.72} & \textbf{0.30}&   \textbf{1.31} &  8.5M         \\\bottomrule
    \end{tabular}
    }
    \vspace{-0.5cm}
\end{table}

\subsubsection{Speech Reconstruction for Chinese}
To explore the effectiveness of our proposed architecture, we further perform speech reconstruction in Chinese. For the speaker-dependent case, we keep the same training and test splits used in CSSMCM ~\cite{zhao2019cascade} for lip reading; for the speaker-independent case, $S1$ (male) and $S6$ (female) are used for testing and the remaining speakers are used for training and validation.

In Table ~\ref{tab:cmlr-spk-dep}, only LipSound2 results are reported since we make a first attempt at tackling speech reconstruction in Chinese. After checking the generated audio samples, we find that, besides the confusion on segments, there are some tone errors. One of the reasons is that Chinese is a tonal language in which lexical tones play an important role for semantic discrimination. The fundamental frequency (F0) which is produced by the vibration of vocal cords is not visible in the input videos (face area), and it is reported that the visual features have a weak correlation to F0~\cite{cornu2015reconstructing}. Another reason is that the VoxCeleb2 dataset mainly consists of non-tonal languages, e.g. British English, American English and German, which makes the pre-training pay little attention to tone production.

\begin{table}[th]
  \setlength{\abovecaptionskip}{0.1cm}
  \setlength{\belowcaptionskip}{-0.1cm}
  \caption{Speech reconstruction results for Chinese on CMLR datasets.}
  \label{tab:cmlr-spk-dep}
  \centering
  \resizebox{7.5cm}{!}{
    \begin{tabular}{ccccc}
    \toprule
                  & \multicolumn{2}{c}{Speaker-dependent} & \multicolumn{2}{c}{Speaker-independent} \\ \hline
    Model         & ESTOI        & PESQ       & ESTOI          & PESQ          \\ \hline
    LipSound2          &\textbf{0.36}        &       \textbf{1.43} & \textbf{0.28}& \textbf{1.21}             \\ \bottomrule
    \end{tabular}
    }
    \vspace{-0.5cm}
\end{table}

\begin{figure}[H]
  \setlength{\abovecaptionskip}{0cm}
  \setlength{\belowcaptionskip}{-0cm}
  \centering
  \includegraphics[width=7cm]{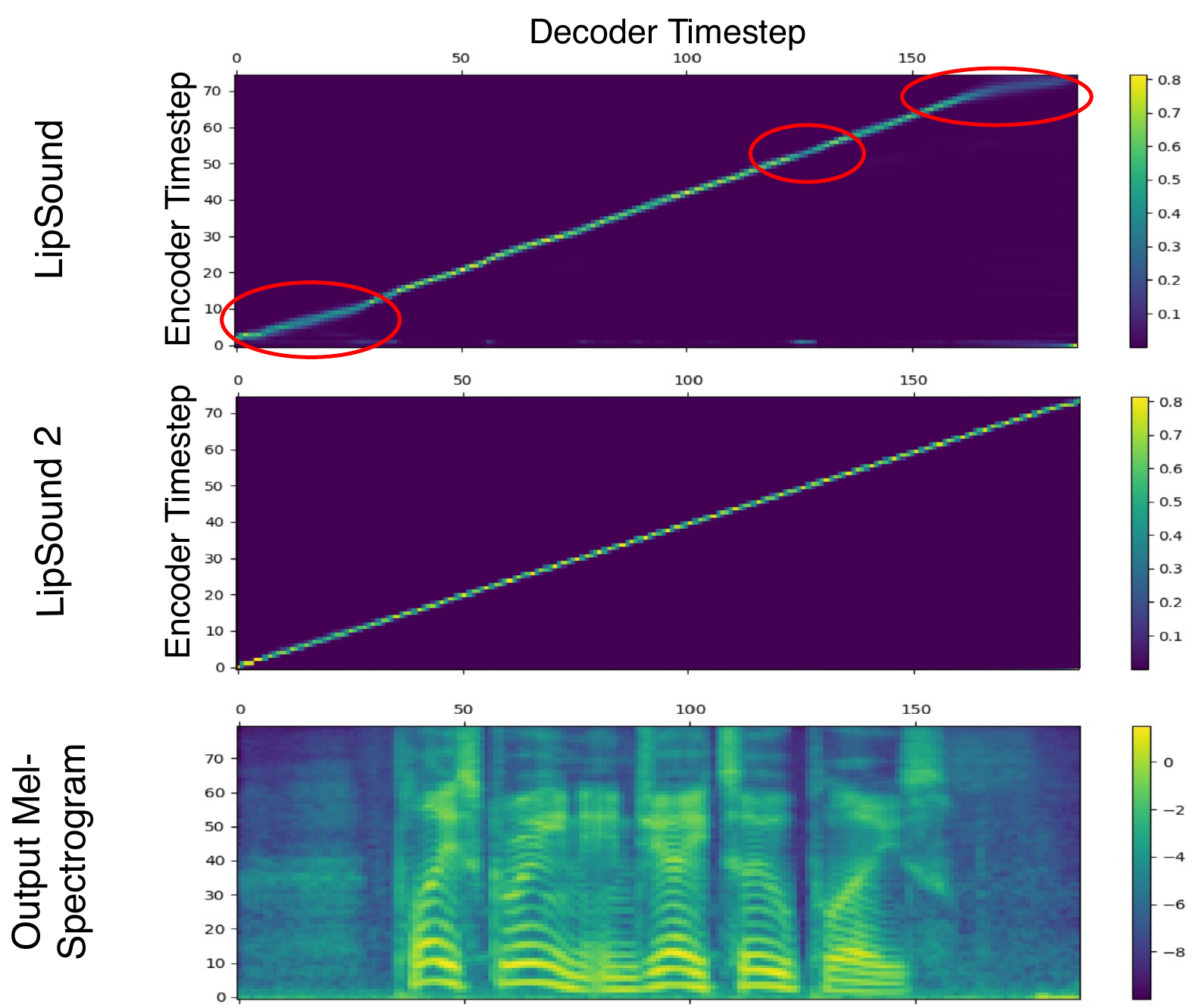}  
  \caption{Attention alignment comparison on GRID dataset.}
  \label{fig:alignment}
\end{figure}

\subsubsection{Attention Alignment}

we compare the attention alignments learned by LipSound~\cite{qu2019lipsound} which is only trained on the GRID dataset and LipSound2 (this paper). As shown in Fig.~\ref{fig:alignment}, the LipSound attention weights are fuzzy at non-verbal areas and at short pauses between words, which may mislead the decoder into focusing on irrelevant encoder timesteps, whereas the attention weights learned by LipSound2 are intensive and more robust to silence or short pauses.

\subsection{Lip Reading Results}
Different from conventional methods which directly transform videos into text,
we perform lip reading experiments in two steps, i.e. video-to-wav and wav-to-text.

\subsubsection{Lip Reading Results for English}

we follow the same splits as previous works for training and test on GRID~\cite{assael2016lipnet} and TCD-TIMIT~\cite{thangthai2018comparing} datasets. The comparison with related results are listed in Table ~\ref{tab:lip reading-grid-timit}. We report the WER of GRID and TCD-TIMIT audio test sets on pre-trained acoustic models (Audio Gold Standard) and the results fine-tuned on the training audio samples (+Fine-Tuning), which is treated as the upper boundary of lip reading.

Our LipSound2 model achieves state-of-the-art performance on both GRID and TCD-TIMIT datasets. Fine-tuning the acoustic model pretrained on 960h LibriSpeech with generated audios can not only significantly boost the model performance but also accelerate training time.

Further improvement can be achieved when an external language model is integrated. The benefit from the language model on the GRID dataset is not as much as on TCD-TIMIT, since the sentence structure in GRID is designed by an artificial grammar. The language model can only help to correct misspelled words but cannot contribute grammatically or semantically.

\begin{table}[th]
  \setlength{\abovecaptionskip}{0cm}
  \setlength{\belowcaptionskip}{-0.1cm}
  \caption{lip reading results on GRID and TCD-TIMIT dataset on WER. Spk-Dep: Speaker-Dependent. Spk-Indep: Speaker-Independent. LM: Language Model.}
  \label{tab:lip reading-grid-timit}
  \centering
  \resizebox{9cm}{!}{
    \begin{tabular}{ccccc}
    \toprule
              & \multicolumn{2}{c}{GRID}                & \multicolumn{2}{c}{TCD-TIMIT}           \\ \hline
Model         & Spk-Dep & Spk-Indep & Spk-Dep & Spk-Indep \\ \hline
Audio Gold Standard    & 22.36            & 21.88              & 15.86             & 15.21 \\
\hspace{1cm}  +Fine-tuning & 0.15            & 0.35               & 5.42             & 6.73 \\ \hline
LipNet~\cite{assael2016lipnet}    & 5.6            & 13.6               & -             & - \\
LipNet+LM~\cite{assael2016lipnet}    & 4.8             & 11.4               & -             & - \\
PCPG+LM~\cite{luo2020pseudo}    & -             & 11.2               & -             & - \\
TVSR-Net~\cite{yang2020speaker}    & -           & 9.1              & -             & - \\

WAS~\cite{chung2017lip}    & 3.0             & -               & -             & - \\
LCANet\cite{xu2018lcanet}    & 2.9             & -               & -             & - \\
DualLip~\cite{chen2020duallip}    & 2.7           & -               & -             & - \\
LipSound~\cite{qu2019lipsound}    & 2.5           & -               & -             & - \\

CD-DNN~\cite{thangthai2018comparing}    & -           & -               & 51.26             & 57.03 \\

MobiLipNetV2~\cite{koumparoulis2019mobilipnet}       & -            & -               & -             & 53.01               \\ \hline
LipSound2          & 1.9        &       7.3 & 41.37& 46.29             \\
LipSound2 + LM      & \textbf{1.5}         &     \textbf{6.4}  & \textbf{39.77}&  \textbf{43.53}          \\ \bottomrule
    \end{tabular}
    }
\end{table}

\subsubsection{Lip Reading Results for Chinese}
we also explore lip reading performance in Chinese, as shown in Table~\ref{tab:lip reading-cmlr}. Audio Gold Standard is directly evaluating the CMLR test set on a pre-trained acoustic model trained on 1000h AISHELL2 dataset. After fine-tuning with CMLR training audios, we get 3.88\% CER and 4.89\% CER for speaker-dependent and -independent cases respectively.

In comparison to other work, our LipSound2 model achieves better results. CER further drops when decoding with an external language model. Besides, we build a new baseline for CMLR in speaker-independent settings.

\begin{table}[th]
  \setlength{\abovecaptionskip}{0cm}
  \setlength{\belowcaptionskip}{-0.1cm}
  \caption{lip reading results for Chinese on CMLR datasets. CER: character error rate.}
  \label{tab:lip reading-cmlr}
  \centering
  \resizebox{6cm}{!}{
    \begin{tabular}{ccccc}
    \toprule
Model         & Spk-dep & Spk-indep \\ \hline
Audio Gold Standard & 19.25   & 16.2             \\ 
\hspace{1cm}  +Fine-tuning & 3.88   & 4.89             \\ \hline
WAS~\cite{chung2017lip}    & 38.93   & -               \\
CSSMCM~\cite{zhao2019cascade}   & 32.48   & -               \\
LIBS~\cite{zhao2020hearing}  & 31.27   & -               \\\hline
LipSound2          & 25.03        &  36.56                   \\
LipSound2 + LM      & \textbf{22.93}         & \textbf{33.44}              \\ \bottomrule
    \end{tabular}
    }
\vspace{-0.5cm}
\end{table}

\section{Discussion}
\label{discussion}

Although the proposed LipSound2 model pre-trained on a large-scale dataset achieves considerable performance on both speech reconstruction and lip reading tasks, it still generates error speech due to the visual similarity on pronunciation, for example, 'pill' is easy to be misrecognized as 'bill' in English and 'ji zhi' is mistaken as 'qi zhi' in Chinese. In addition, our model can generate quite similar voices as the ground truth in speaker-dependent settings, while the model is inclined to predict a voice existing in training set sometimes in speaker-independent cases. For
details and a demonstrations we refer also to the demo video on the project website~\footnote{https://leyuanqu.github.io/LipSound2/}. How to stop the fine-tuning procedure at the appropriate time and avoid the model overfitting on downstream tasks is an important direction for future research, since the MSE loss always declines when using teacher forcing during training, which hardly indicates whether the model is overfitting or not. Besides, a possible solution could be using voice embeddings as additional inputs that can efficiently help models learn speaker identity information, as we found in our previous work~\cite{qu2020multimodal}. 
\vspace{-0.1cm}

\section{Conclusion}
\label{conclusion}

In this paper, we have proposed LipSound2 which directly predicts speech representations from raw pixels. We investigated the effectiveness of self-supervised pre-training for speech reconstruction on large-scale vocabulary datasets, particularly for speaker-independent settings. Moreover, state-of-the-art results are achieved by fine-tuning the produced audios on a well pretrained speech recognition model for both English and Chinese lip reading experiments, since our two-step method benefits not only from the large-scale crossmodal supervision which enables the model to learn more robust representations and more different content information, but also from the advanced speech recognition architecture (acoustic and language models) which is pre-trained on abundant labeled data.

Although we have made great progress on speech reconstruction in controlled environments, there is still a significant gap to the requirements of real-world scenarios. Future work will focus on more realistic configuration, such as the variety of light conditions, moving head poses and different background environments. Moreover, the current lip reading experiments are separately conducted in two steps in which the error generated in the first step (video-to-wav) will be propagated to the second step (wav-to-text). How to jointly train the two tasks in an end-to-end fashion could be another direction. Besides, we are also interested in integrating our LipSound2 model into active speaker detection, speech enhancement and speech separation tasks to boost the performance of speech recognition systems in human-robot interaction. 


\section*{Acknowledgement} 
The authors gratefully acknowledge partial support from the China Scholarship Council (CSC) and from the German Research Foundation DFG under project CML (TRR 169).

\bibliographystyle{IEEEtran}{
\bibliography{lipsound2}}
\vspace{-30 pt}
\begin{IEEEbiography}[{\includegraphics[height=1.25in,clip,keepaspectratio]{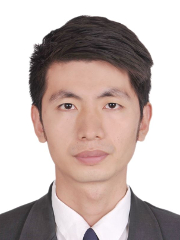}}]{Leyuan Qu} received the M.Sc degree in computer science from Beijing Language and Culture University, Beijing, China, in 2017. He is currently a Ph.D. student in the Department of Informatics, University of Hamburg, Hamburg, Germany. His main research interests include robust speech recognition, audio-visual speech recognition \& speech enhancement \& speech separation, lip reading and self-supervised learning.
\end{IEEEbiography}

\vspace{-30 pt}

\begin{IEEEbiography}[{\includegraphics[height=1.25in,clip,keepaspectratio]{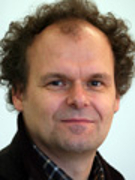}}]{Cornelius Weber} received the Diploma degree in physics, from the University of Bielefeld, Bielefeld, Germany, and the Ph.D. degree in computer science with the Technische Universität Berlin, Berlin, Germany, in 2000. He is a Laboratory Manager with the Knowledge Technology Group, Universität Hamburg, Hamburg, Germany. He was a Post-Doctoral Fellow of Brain and Cognitive Sciences with the University of Rochester, Rochester, NY, USA. From 2002 to 2005, he was a Research Scientist of Hybrid Intelligent Systems with the University of Sunderland, Sunderland, U.K. He was a Junior Fellow with the Frankfurt Institute for Advanced Studies, Frankfurt am Main, Germany, until 2010. His current research interests include computational neuroscience with a focus on vision, unsupervised learning, and reinforcement learning.

\end{IEEEbiography}

\vspace{-30 pt}

\begin{IEEEbiography}[{\includegraphics[height=1.25in,clip,keepaspectratio]{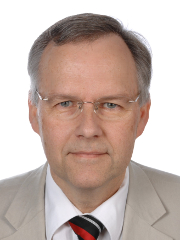}}]{Stefan Wermter} is Full Professor at the University of Hamburg, Germany, and Director of the Knowledge Technology Institute in the Dept. of Informatics. His main research interests are in the fields of neural networks, hybrid knowledge technology, cognitive robotics, and human-robot interaction. He has been an associate editor of the journals ‘Transactions on Neural Networks and Learning Systems’, and is an associate editor of ‘Connection Science’ and ‘International Journal for Hybrid Intelligent Systems’ and he is on the editorial board of the journals ‘Cognitive Systems Research’, ‘Cognitive Computation’ and ‘Journal of Computational Intelligence’. Currently, he is co-coordinator of the international collaborative research centre on Crossmodal Learning (TRR-169) and coordinator of the European Training Network SECURE on safety for cognitive robots. In 2019, he has been elected as the President for the European Neural Network Society 2020-2022.
\end{IEEEbiography}

\end{document}